\documentstyle[geodyn]{article}

\lefthead{VISHNIAC}

\righthead{GALACTIC AND ACCRETION DISK DYNAMOS}

\authoraddr{E.T. Vishniac,
Department of Physics and Astronomy,
Johns Hopkins University,
3400 N. Charles St.,
Baltimore MD 21218.
(e-mail: ethan@pha.jhu.edu)}

\begin{document}

\title{Galactic and Accretion Disk Dynamos}

\author{Ethan T. Vishniac, \altaffilmark{1} \altaffilmark{2}}

\affil{Department of Physics and Astronomy, Johns Hopkins University, Baltimore, Maryland}

\altaffiltext{1}{Visiting Professor 1997-98 in the Department of Physics,
Massachusetts Institute of Technology, Cambridge, Massachusetts}

\altaffiltext{2}{Department of Astronomy, The University of Texas, Austin, Texas}

\begin{abstract}
Dynamos in astrophysical disks are usually explained in terms of the
standard alpha-omega mean field dynamo model where the local helicity
generates a radial field component from an azimuthal field. The subsequent
shearing of the radial field gives rise to exponentially growing dynamo 
modes.  There are several problems with this model.  The exponentiation
time for the galactic dynamo is hard to calculate, but is probably 
uncomfortably long.  Moreover, numerical simulations of magnetic fields in
shearing flows indicate that the presence of a dynamo does not depend on a
non-zero average helicity.  However, these difficulties can be overcome
by including a fluctuating helicity driven by
hydrodynamic or magnetic instabilities.  Unlike traditional disk dynamo models,
this `incoherent' dynamo does not depend on the presence of systematic fluid helicity
or any kind of vertical symmetry breaking.  It 
will depend on geometry, in the sense that the dynamo growth rate becomes smaller for
very thin disks, in agreement with constraints taken from
the study of X-ray novae.  In this picture the galactic dynamo will
operate efficiently, but the resulting field will have a radial coherence
length which is a fraction of the galactic radius.  
\end{abstract}

\begin{article}
\section{CONTEXT}
 
The traditional focus of astrophysical dynamo theory has been
on stars, where spherical symmetry is a reasonable first
approximation, and the inward pull of gravity is balanced
by the radial pressure gradient.   In spite of the eponymous
role of stars in astrophysics, this ignores the importance of
magnetic fields in disks, where gravity is balanced by
centrifugal forces.  This traditional bias can be
explained by the fact that we can observe the magnetic
fields of at least one star in some detail, whereas the
magnetic field of the Galactic disk presents itself
as a bewildering mixture of structure on a range of
scales.  However, recent years have witnessed an
accumulation of data concerning the structure of the
magnetic field in our galaxy, and in external galaxies.
In addition, it has become clear that magnetic fields
play a critical dynamical role in accretion
disks of all sizes, including some of the most luminous
objects in the universe.  Here I will summarize recent 
theoretical progress in understanding disk dynamos.
In an unexpected twist, we will see that the role of
global helicity in magnetic field generation may be 
small.

We start by considering the context of disk dynamos.
Astrophysical disks can be divided into two general categories,
galactic disks and accretion disks.  The latter
category includes disks around the supermassive black holes,
as in active galactic nuclei (AGN), and stellar disks surrounding
protostars or members of binary star systems.  Although the
physical conditions in these disks span an enormous
range we will restrict ourselves in only two ways.
{}First, we consider disks that are sufficiently ionized
that ohmic dissipation is negligible on the scale of the
disk thickness.  This may exclude parts of protostellar
disks, and some regions in the interstellar medium (ISM).
Second, we restrict ourselves to disks that are 
primarily supported by rotation, and are consequently
geometrically thin.  There are cases where a large
fraction of the disk support comes from radial pressure
gradients or from the magnetic field of the accreting
object.  We ignore these cases because of the complicated
physics involved, not because we believe them to be
unimportant.

With this in mind, we can summarize the differences between
galactic and accretion disks in the following manner.
Galactic disks are confined vertically by the gravity
of the disk and halo acting together.  The typical
galactic disk has an angular velocity $\Omega\propto r^{-1}$
over a broad range in radii.  The gaseous disk itself is composed
of a heterogeneous interstellar medium with a complicated
history.  It is typically marginally stable against local
gravitational collapse.  The number of dynamical
time scales since the formation of the galaxy is limited,
and it is unclear whether or not the magnetic field in
the disk can be regarded as having reached a stationary
state, or whether initial conditions might be important
in understanding its structure.  Finally, the orbital
period is the longest natural time scale in these
systems, followed by the dynamical time scale of local
random motions within the disk, followed by the
particle collision time in the gas. This is, in turn,
greater than plasma time scales, such as the inverse
of the ion cyclotron frequency or the inverse of the
plasma frequency.  Treating the gas in a galactic
disk as a fluid is clearly a dangerous approximation,
both because of its complicated substructure, and because
the hydrodynamic approximation is unlikely to be
accurate even within a relatively homogeneous
volume of the ISM.  Most of the volume of the gaseous
disk is occupied by gas that is sufficiently ionized
that ohmic dissipation is negligible on disk scales.

In contrast, vertical confinement in accretion disks is
supplied by the vertical component of the gravity of the
central object.  This leads to an orbital frequency
$\Omega\propto r^{-3/2}$ and, through the condition
of hydrostatic equilibrium a thickness $H\propto c_s/\Omega$,
where $c_s$ is the local sound speed.  Accretion disks
are relatively homogeneous, in the sense that the vertical sound 
crossing times are short and pressure equilibrium is 
a good approximation. In the absence of magnetic fields
accretion disks are stable, although strongly unstable when
they are present.  Their age is greater than all other
relevant time scales, so initial conditions can be ignored.
The time it takes material to spiral inward to the central
object is greater than the time scale for local thermal
equilibrium, which is greater than an orbital period.
Inverse plasma frequencies are typically much less than
an orbital period, and usually much greater than the
mean free collision time for a particle.  Accretion disks
are good fluids, although accretion disk coronae are not.
Accretion disks are not always good conductors, but
the exceptions are cold and difficult to observe.

\section{CLUES AND CONSTRAINTS}

Our knowledge of astrophysical magnetic fields is never
as complete as we would like.  For galactic magnetic
fields we have a variety of diagnostics which tells us
about the current state of the field.  Direct observations
of evolutionary effects are, of course, impossible.  These
diagnostics include the intensity and polarization of
synchrotron radiation, the polarization of starlight,
the polarization of infrared dust emission, Faraday
rotation, and Zeeman splitting.  (For a general review
of galactic magnetic fields see Zweibel and
Heiles 1997; Vall\'ee 1997).  It is important to note that each of these
diagnostics involves other quantities, for example electron
density or the physical properties of interstellar dust
grains, for which we have only rough estimates.  In addition,
the direction of the magnetic field can be derived only
from Faraday rotation, and only for the component along the
line of sight.  

Keeping these uncertainties in mind, we note that a rough
concurrence among these methods allows us to conclude that
the mean value of the magnetic field in the disk is
approximately $10^{-5.5}$ Gauss, with comparable power in
the large scale and `random' (i.e. small scale) components.
The large scale field is approximately aligned with the
azimuthal direction, but tilted somewhat towards the
direction of the local spiral arms.  The number of large
scale field reversals in the disk is unknown, but cannot
be very large, since observations of Faraday rotation
tend to give be consistent with a large scale field coherence
length of at least several hundreds of parsecs.

Models of galactic magnetic field generation usually assign
a rather large role to the galactic shear.  We note that
this is about $10^{-15}\hbox{sec}^{-1}$ at our position in
the Galaxy.  Given a galactic disk age of $\sim10^{10}$ years
this gives a maximum growth of roughly $300$ e-foldings.
There are various suggestions for modifying fundamental physics
in order to obtain a large scale primordial magnetic field,
but these proposals are all highly speculative.  Simply
positing a primordial field as an initial condition poses
severe problems for the successful standard cosmological model.
If we restrict ourselves to magnetic fields generated by
the stresses that accompany the formation of a galactic
disk, then we obtain large scale seed fields in the 
range $10^{-18}$
to $10^{-19}$ Gauss (Lazarian 1992; Kulsrud, Cen, Ostriker and Ryu 1997)
by invoking the Biermann battery in a realistic protogalaxy 
(Biermann 1950; for an exposition in English see Kemp 1982).
This implies about $30$ e-foldings
of growth up to the present day, or a galactic dynamo
growth rate which is no less than ten percent of the local
shear rate.  Since the current epoch in the history of
our galaxy is unlikely to be special, in the sense that
the magnetic field is unlikely to have just reached
equipartition with the gas pressure, we would prefer a
dynamo growth rate comfortably above this minimum.

{}For accretion disks we face a major observational difficulty.
The magnetic field inside an accretion disk is
completely unobservable.  However, there are indirect constraints
on the magnetic field strength.  The luminosity of
an accretion disk depends on the mass transport through the
disk, and indirectly on the average radial velocity of the
disk material.  This is related to the dimensionless `viscosity' $\alpha$
by
\begin{equation}
V_r\approx \alpha{c_s^2\over r\Omega}.
\end{equation}
When a magnetic field is present, local instabilities in the field
(see below) imply
\begin{equation}
\alpha\approx{B_rB_\theta\over 4\pi P}\propto {V_A^2\over c_s^2}.
\end{equation}
In other words, the efficiency of radial mass transport is a measure
of the ratio of magnetic pressure to gas pressure in an accretion disk.

{}For stationary systems this does not allow us to constrain the mean
magnetic field, but the evolution of time varying systems is sensitive
to the actual value of $\alpha$.  In particular, a variety of systems,
including dwarf novae and X-ray novae, undergo recurrent transitions  
between hot, ionized, luminous states and cold, mostly neutral
quiescent states.  The luminous outburst state is marked by a relatively
high mass flux through the disk while the quiescent state transfers
little mass through the disk.  Consequently, each system undergoes
a thermal limit cycle, in which material accumulated near the
outer edge of the disk during quiescence is spread through the disk,
and onto the central object, during an outburst (for a review
see Cannizzo 1993).
Typical bright outbursts are marked by a fast rise
and exponential decay.  The rise marks appearance and spread of the
hot state, typically starting far from the central object.  The
decay corresponds to the reappearance of the cold state, typically
near the outer edge, and the subsequent progress of a cooling front
to small radii.   The duration of the outburst is sensitive to
the rate at which a significant fraction of the total disk mass can be 
deposited on the central star, and therefore is a direct measure 
of $\alpha_{hot}$, the average value of $\alpha$ in the hot state.  
Conversely, the duration of a quiescent phase is a measure of how
much mass can be accumulated without forcing the disk into outburst,
and is therefore a measure of $\alpha_{cold}$.  Finally, the shape
of the luminosity decay at the end of an outburst is a measure of
how the cooling front velocity depends on radius. All of this
data can be fit by taking
\begin{equation}
\alpha\sim 35\left({c_s\over r\Omega}\right)^{3/2},
\label{alpha2}
\end{equation}
which also fits the difference in time scales between black hole candidate
systems, with a central mass $\sim 7 M_{\sun}$, and white dwarf systems
(Cannizzo, Chen, and Livio 1995; Vishniac and Wheeler 1996).
The ratio $c_s/(r\Omega)$ is not necessarily a sign that the orbital velocity
of the disk material is directly connected to the dynamo rate.  This is
also the ratio of the disk height to radius and may have a purely
geometric origin.

\section{LOCAL MAGNETOHYDRODYNAMIC INSTABILITIES IN DISKS}

In a purely hydrodynamic disk, i.e. when no magnetic field is present, 
there are no local instabilities aside from those induced by self-gravity
or tidal effects from a companion.  This encourages us to treat the
evolution of a magnetic field in a smooth background.  The dispersion
relation is
\begin{eqnarray}
  \biggl[1- & {(x^2+3x_A^2)\over(x^2-x_A^2)^2(1+\kappa^2)}
+{9\over2}{\kappa^2x_A^2\over(1+\kappa^2)x^2(x^2-x_A^2)}\biggr]u_r=
\nonumber\\
&={9\over4}{\kappa^2\over1+\kappa^2}\partial_x^2u_r
+{9\over2}{\kappa^2\over1+\kappa^2}{x_A^2\over(x^2-x_A^2)x}\partial_xu_r,
\end{eqnarray}
for radial scales $\ll r$ and ignoring the vertical structure of the
disk (Vishniac and Diamond 1992; Matsumoto and Tajima 1995).  
In this equation
\begin{equation}
x\equiv{\bar\omega\over\Omega}={\omega\over\Omega}+k_{\theta}r,
\end{equation}
\begin{equation}
x_A\equiv{\omega_A\over\Omega}={\vec k\cdot\vec B\over(4\pi\rho)^{1/2}\Omega},
\end{equation}
\begin{equation}
\kappa\equiv{k_{\theta}\over k_z},
\end{equation}
and $u_r$ is the radial velocity perturbation.  The frequency $\bar\omega$
is the frequency measured by an observer rotating with the local fluid 
speed and $\omega$ is the frequency measured by an external observer.
Since $\omega$ is a global quantity, while the dynamics of the perturbation
are determined by $\bar\omega$, which is a function of radius, the
radial dependence cannot be generally assumed to be described by
some radial wavenumber.  Here we have taken advantage of the radial
dependence of $\bar\omega$ to use $x$ as a radial coordinate.

In the axisymmetric limit this expression gives an instability.  It is
less obvious when $k_{\theta}\ne 0$ but this instability is generally
present.  It was first discovered by Velikhov (1959), and independently
by Chandrasekhar (1961), and first applied to accretion disks by
Balbus and Hawley (1991).  Physically it is related to the famous
tethered satellite experiment, except that it works.  If magnetic
field lines in the vertical or azimuthal direction are perturbed
radially, then gas at smaller radii can transfer angular momentum
outward to the slower moving gas on the same field line.  This
works whenever $\Omega$ increases inward while specific angular
momentum increases outward.  In a accretion disk the large scale 
azimuthal field tends to dominate, so the non-axisymmetric case
is the most important.  One additional subtlety is that local
nonaxisymmetric disturbances do not correspond to global linear
modes, and only grow $\sim k_z/k_\theta$ e-foldings before
dissipating, but this is sufficient to ensure local instability in
any practical sense of the phrase.

Our expectation, based on this linear dispersion relation, is that the
dominant modes will have growth rates comparable to $\Omega$, and
wavelengths of roughly $V_A/\Omega$ in all directions.

\section{NUMERICAL SIMULATIONS}

Linear theory gives us some understanding of the driving force behind the 
transition to turbulence, and consequently a set of dimensional estimates
for the nature of the turbulent regime.  However, any hope of obtaining
a quantitative understanding of real systems has to rest with numerical 
simulations.  A number of groups have attempted simulations of
the growth of the Balbus-Hawley instability in accretion disks (see, for
example Brandenburg, Nordlund, Stein, and Torkelsson 1996; Hawley, Gammie,
and Balbus 1996; Stone, Hawley, Gammie, and Balbus 1996).
While these simulations have not completely overcome the technical
difficulties involved in following MHD turbulence over a broad dynamical
range, they do show some common results which we can take as a guide
in considering real accretion disks.  
Since the nature of the simulations may play a large role in the results,
we need to consider their common elements.  First, in order
to reduce the problem to a manageable size, the disk is idealized
as a fluid in a shearing flow, with a scale height 
which is comparable to its radial extent.  Rather than simulate
an entire annulus the usual procedure is to make the box periodic in
the azimuthal direction, with a total length which is typically about
$2\pi$ vertical scale heights.  (Although there have been simulations with 
azimuthal lengths up to four times longer.)  

What do the results look like?  
{}First, naive expectations based on linear theory appear to be correct.
There is a transition to turbulence, with the scales expected from the
linear analysis.  The resulting eddies are  moderately anisotropic with 
$\lambda_\theta>\lambda_r>\lambda_z$.

Second, the evolution of the magnetic field typically has two phases.
At first the magnetic field strength grows exponentially, with a rate
$\sim\Omega$.  However, this growth involves short wavelength components
of the field.  When this phase saturates, a slower growth appears,
in which the large scale field components acquire a substantial fraction
of the total magnetic energy.  This latter phase frequently includes large
scale field reversals, with a frequency which is roughly comparable to
the growth rate of the large scale field.

Third, at saturation the field typically shows $\langle v^2\rangle$ a 
fraction of $V_A^2$, which is in turn a large fraction of $c_s^2$.  We
expect $\alpha$ to scale with $(V_A/c_s)^2$, but in practice it remains
small, typically less than a percent.  However, the value of $\alpha$
varies from one simulation to another and appears to increase with
increasing numerical resolution.  It is plausible to suppose that
for realistic Reynolds numbers $\alpha$ would reach reasonable 
values, although this involves a considerable amount of extrapolation.

{}Finally, one of the more striking features of this work is that the results are
not qualitatively different for simulations which include vertical
stratification and those that simply confine the fluid in a box
with periodic vertical boundary conditions.  In other words, vertical
symmetry breaking does not play an important role in the dynamo
present in these simulations.

\section{DYNAMO THEORY}

\subsection{Conventional $\alpha-\Omega$ Dynamos}

What generates the large scale field in the simulations, or, for that matter,
in astrophysical disks?  The usual answer is to appeal to mean field
dynamo theory.  In the context of strongly shearing astrophysical disks,
the evolution equations for the large scale field can be written in
a simplified form, i.e.
\begin{equation}
\partial_t B_r\approx -\partial_z(\alpha_{\theta\theta}B_\theta)
+\partial_z D_T\partial_z B_r,
\label{br1}
\end{equation}
and
\begin{equation}
\partial_t B_\theta\approx -{3\over2}\Omega B_r
+\partial_z D_T\partial_z B_\theta,
\label{bt1}
\end{equation}
where $D_T$ is the turbulent diffusivity and 
\begin{equation}
\alpha_{\theta\theta}=
\langle v_z\partial_\theta v_r-v_r\partial_\theta v_z\rangle\tau.
\end{equation}
Here $\tau$ is the velocity correlation time.  This formulation of
mean field dynamo theory is referred to as the `$\alpha-\Omega$ dynamo,
since the radial field is generated from the azimuthal field by helicity
and the cycle is closed by the shearing of the radial field to create
azimuthal field.

In order have a non-zero $\alpha_{\theta\theta}$ we need to have
some systematic violation of symmetry with respect to the $\hat z$
direction.  The same is also true for radial and azimuthal motions,
but coriolis forces can be relied upon to generate correlations
between motions and gradients in these two directions.  Vertical
symmetry breaking requires the presence of vertical stratification.
However, as we saw in the last section, this does not play a crucial
role in the simulations.   Whatever dynamo is operating in them
is indifferent to whether or not $\alpha_{\theta\theta}=0$.

Notwithstanding this point, there have been several attempts to
derive a dynamo theory for accretion disks using magnetic field
buoyancy, or more specifically, the Parker instability (see, for 
example Tout and Pringle 1992).  
These models all face a basic theoretical problem.  The
growth rate for the Parker instability is of order $(V_A/c_s)\Omega$
with the fastest growing modes having azimuthal wavelengths similar
to the pressure scale height, or $c_s/\Omega$.  Shearing constraints
imply that the corresponding radial wavelengths are of order
$V_A/\Omega$, which is also the typical radial scale for the Balbus-Hawley
instability.  Consequently, rising and falling sections of the magnetic
field are mixed at a rate $\sim \Omega$.  Unless the magnetic field
is already strong (i.e. $V_A \sim c_s$) this is much faster than the
growth rate of the Parker instability.  In fact, numerical simulations
with vertical stratification show little sign of the Parker instability,
even when $V_A$ is large.

\subsection{Incoherent and Chaotic Dynamos}

What are the alternatives to the standard $\alpha-\Omega$
dynamo?  One idea is that the magnetic field is sustained
through a local, chaotic process, in which local field stretching
amplifies the field up to equipartition with the ambient
pressure.  This picture was originally suggested by Batchelor
(1950), although the first detailed treatment is due to 
Kazantsev (1967).  It can be rigorously justified only in
the limit of a weak magnetic field, which is never the case
when the turbulence itself is driven by the field.  In any
case, if we accept this possibility in accretion disks then
the large scale field would then be explained as the
result of some sort of inverse cascade within a turbulent
fluid.  This model is not
consistent with accretion disk phenomenology, in particular the
thermal limit cycle and the decay from outburst of dwarf novae and
soft X-ray transients mentioned above. It is also unclear
why the very largest scales, with wavelengths equal to several
eddy scales, always end up with a significant fraction of the total power.

An alternative explanation is that the large scale field is
generated by an extension of the $\alpha-\Omega$ dynamo developed
by Vishniac and Brandenburg (1997) called the `incoherent dynamo'.
In the simplest version of this model the vertical symmetry is
assumed to be unbroken, so that $\langle\alpha_{\theta\theta}\rangle=0$.
However, at any moment a magnetic domain containing $N$ eddies will
have a helicity of
\begin{equation}
\alpha_{\theta\theta}\sim{\alpha_{\theta\theta,E}\over N^{1/2}},
\end{equation}
where $\alpha_{\theta\theta,E}$ is the helicity associated with a single
eddy, which is comparable to the turbulent eddy velocity, $V_T$.  
In this case equation (\ref{br1}) can be
written as a stochastic equation.  It is also helpful to rewrite it
in terms of the evolution of $\langle B_r^2\rangle $ or
\begin{equation} 
\partial_t\langle B_r^2\rangle=
2\langle\left[\partial_z(\alpha_{\theta\theta}B_\theta)\right]^2\rangle
\tau-2D_T\langle (\partial_z B_r)^2\rangle.
\label{br2}
\end{equation} 
(Here the brackets denote only spatial averaging.)  Combining equations
(\ref{bt1}) and (\ref{br2}) we can estimate the incoherent dynamo
growth rate as
\begin{equation}
\gamma\approx\left({\langle\alpha_{\theta\theta,E}^2\rangle\tau\Omega^2
\over L_z^2 N}\right)^{1/3},
\end{equation} 
where $L_z$ is the vertical height of a magnetic domain.
This growth is a combination of random walk in $B_r$, driven
by fluctuations in $B_\theta$, and the shearing of $B_r$.
The fact that it gives exponential growth results from a tendency
for the distribution of $B_r/B_\theta$ to be biased towards negative
numbers.  When this ratio becomes sufficiently positive the field
undergoes a sudden reversal and $B_{\theta}$ switches sign.
Typically non-axisymmetric domains are sheared out faster than 
they can grow, so the usual expression for $N$ in isotropic
turbulence will be
\begin{equation}
N\approx {L_z L_r 2\pi r\over\lambda_T^3}.
\end{equation}

Incoherent dynamos are intrinsically noisy.  In addition to the 
field fluctuations on eddy scales, the large scale field will
undergo spontaneous field reversals with a frequency not far
below the dynamo growth rate.  Furthermore, the coupling between
different domain scales implies that there is constant `crosstalk'
between different Fourier modes of the large scale magnetic field.
Consequently, there are no well-defined 
linear eigenfunctions of this dynamo.
Since individual annuli switch polarity on a regular basis, there
seems little chance that the disk magnetic field will become
uniform on radial scales larger than a few disk scale
heights.  Furthermore, this will reduce the strength of any large
scale poloidal field produced via magnetic buoyancy.  Different
disk annuli will contribute randomly to any global field.

Finally, if we compare the growth rate $\gamma$ to the dissipation
rate, $\sim V_T^2\tau/L_z^2$, we see that the largest vertical scale
domains will accumulate most of the energy.  (This line of reasoning
can't be used to argue for larger radial scales since extending magnetic
domains radially will lower the growth rate while leaving the dissipation
rate unchanged.)

\section{THE INCOHERENT DYNAMO IN ASTROPHYSICAL DISKS}

\subsection{Accretion Disks}

If we wish to apply the incoherent dynamo to accretion disks then
the obvious source of small scale turbulence is the Balbus-Hawley
instability.  Aside from the point that this is the
only source of turbulence which is guaranteed to accompany a
successful dynamo, only very strong convection is likely to
survive the turbulent mixing caused by magnetic field instabilities.
In this case we can write the dynamo growth rate as
\begin{equation}
\gamma\sim\left[\left({V_A\over c_s}\right)^5{H\over r}G(\beta)\right]^{1/3},
\end{equation}
where $G(\beta)$ describes the saturation of this mechanism as the
ratio of magnetic to ambient pressure ($\beta^{-1}$) approaches unity.
Here I have assumed that the magnetic domain is about as thick and wide
as a disk vertical scale height.  The dissipation rate is proportional
to $\beta^{-1}\Omega$.  This implies that the growth rate of the dynamo
scales as the magnetic field strength to the $5/3$, while the dissipation
rate scales as the magnetic field strength squared.  Consequently, the
saturated state will be sensitive to other aspects of the model, including
numerical viscosity in the computer simulations.

We can get a sense of how this works for accretion disks by assuming
\begin{equation}
\alpha={1\over3}\left({V_A\over c_s}\right)^2,
\end{equation}
and
\begin{equation}
G=1-{V_A^2\over c_s^2}.
\end{equation}
Both of these are meant to be illustrative rather than serious predictions,
however they have roughly the properties we expect for the exact solution.
The function $G$ should cut off sharply as $V_A\rightarrow c_s$, since
in this limit the Balbus-Hawley instability disappears.  Furthermore,
for $V_A\ll c_s$ we expect $G$ to have a leading order correction term
of order $\beta^{-1}$.  The scaling of $\alpha$ is roughly consistent
with the numerical simulations, but a bit on the high side, reflecting
our expectation that current simulations tend to underestimate its value.
Balancing dynamo growth and turbulent dissipation we find that
\begin{equation}
\alpha={C_0^6\over 3}\left({c_s\over r\Omega}\right)^2(1-3\alpha)^2.
\label{alpha1}
\end{equation}
The value of $C_0$ is difficult to estimate, and in any case is raised
to such a high power that it has to be regarded as essentially a free
parameter.

Applying equation (\ref{alpha1}) to real disks requires us to fit to
phenomenological models of dwarf novae and soft X-ray transients.
If we take $C_0\sim 3$ then we can produce an acceptable fit
to equation (\ref{alpha2}).  In this case we find that for
values of $c_s/(r\Omega)$ between $1$ and $1/4$ the predicted
value of $\alpha$ drops from $0.32$ to $0.29$, i.e. negligibly.
For values of $c_s/(r\Omega)$ more appropriate for dwarf novae 
systems, in the range $0.04$ to $0.025$, $\alpha$ drops to the
range $0.15$ to $0.1$, with a slope with respect to $c_s/(r\Omega$
of $0.75$ to $1$.  Finally, if we take $c_s/(r\Omega)$ down to
one percent, as expect for soft X-ray transients, we 
get $\alpha\approx 0.03$ with a slope of $\sim 5/3$.
These values and slopes are consistent with models of these systems
and with the results of computer simulations.  The extremely weak
response of $\alpha$ to changes in the disk height to radius ratio
when that ratio is not extremely small seems a bit odd.  However,
it is mostly the result of taking $C_0^6/3$ large, which is required
by the thin disk models.  A considerably smaller contribution to
this effect comes from the sharp cutoff in $G$ as $V_A\rightarrow c_s$.  
Both of these effects
are intrinsic to the incoherent dynamo model and would be expected
in any phenomenologically acceptable version of the model.

\subsection{Galactic Disks}

Aside from the differences already noted between galactic and
accretion disks, there is another point which is critical for
any application of the incoherent dynamo to galactic disks.
Since galactic magnetic fields start out weak, the scale of
turbulence due to magnetic instabilities would have been small,
and the incoherent dynamo would have been relatively ineffective.
In order to have a strong dynamo from very early times we need
to appeal to other sources of turbulent motion.  In the case of
a galactic disk, one plausible source would be local gravitational
instabilities.  Another might be violent outflows from
star forming regions.  In either case it is difficult to assign
length scales and velocities from first principles.

Suppose we take the point of view that the kinds of motions
present at early times were not very different from what we
see today.  If we take 
\begin{equation}
V_T\sim 10\hbox{\ km/sec},
\end{equation}
\begin{equation}
L_T\sim 300\hbox{\ parsecs},
\end{equation}
and assume a magnetic field vertical scale of 1 kpc, then we get
a growth rate of
\begin{equation}
\gamma\sim 10^{-16} \hbox{\ sec}^{-1},
\end{equation}
with a slightly smaller dissipation rate.  This estimate is just marginally
fast enough, but ignores factors of order unity, which are bound to
be important in this case.  The only conclusion we can draw from
this exercise is that it is {\it possible} that the incoherent dynamo is
responsible for the growth of large scale galactic fields, but any
real answer will require a firmer understanding of turbulence in the
Galactic disk.  On the other hand, the incoherent dynamo does make
a testable prediction.  Since the growth time is only slightly less than
the reversal time, and since the typical magnetic domain has a radial
extent comparable to the disk thickness, it follows that we should
expect the large scale $B_\theta$ to reverse over radial scales
slightly larger than the disk thickness.  This is consistent with
current observations (see references in Zweibel and Heiles 1997),
but the number of galaxies with observed reversals is still very small.

\section{SUMMARY}

We note several points in conclusion.  First, disk dynamos do not
require an average fluid helicity.  They may require a mean square
helicity, but this is a by-product of turbulence in general.

Second, incoherent dynamo effects match phenomenological constraints
on accretion systems.  They are not inconsistent with numerical simulations,
but are not yet clearly confirmed by such work.  A clear signature
of their presence would be a turn-down in the value of $\alpha$ in
the limit of very long computational boxes.

Third, the incoherent dynamo may be relevant for the rapid growth of
galactic fields.  However, models are sensitive to assumptions
about the properties of turbulence in galactic disks.  The only
clear prediction is that large scale field reversals should be
common on radial scales of a kiloparsec or more.

Fourth, at odds with the general theme of this conference, it is
difficult to find a major role for either fluid or magnetic
helicity in simulations of disk dynamos, or, perhaps, inside real
astrophysical disks.  The interaction of the disk field with
its environment may present a mechanism for the generation of
magnetic helicity by disks (cf. R. Matsumoto's contribution to
this volume).

\acknowledgments
The work presented here was supported in part by 
part NAG5-2773 and NSF grant AST-9318185 (ETV). I
am grateful for a number of helpful discussions with
A. Brandenburg and E. Zweibel as well as the hospitality
of MIT and the CfA for the 1997-98 academic year.



%
%


	%
	%
	%
	%

\end{article}
\newpage

%
%


	%
	%
\end{document}